\documentclass[useAMS,usenatbib]{mn2e} 
\usepackage{graphicx,amsmath}
\usepackage[usenames]{color}
\usepackage{aas_macros}
\title[Analysis of {\it Kepler} B stars]
{Analysis of {\it Kepler} B stars:\\Rotational modulation and Maia variables}
\author[L. A. Balona, A. S. Baran,  J. Daszy\'nska-Daszkiewicz, P. De Cat] 	
{L. A. Balona$^1$, A. S. Baran$^2$, J. Daszy\'nska-Daszkiewicz$^3$, P. De Cat$^4$\\
$^1$South African Astronomical Observatory, P.O. Box 9, Observatory 7935, Cape Town, South Africa\\
$^2$Uniwersytet Pedagogiczny w Krakowie, ul. Podchorazych 2, 30-084, Krak\'{o}w, Poland\\
$^3$Instytut Astronomiczny, Uniwersytet Wroc{\l}awski, Kopernika 11, 51-622
Wroc{\l}aw, Poland\\
$^4$Royal Observatory of Belgium, Ringlaan 3, B-1180 Brussel, Belgium
}

\begin{document}

\date{Accepted .... Received ...}

\pagerange{\pageref{firstpage}--\pageref{lastpage}} \pubyear{2011}

\maketitle

\label{firstpage}

\begin{abstract}
We examine 4-yr almost continuous {\it Kepler} photometry of 115 B stars.
We find that the light curves of 39\,percent of these stars are simply
described by a low-frequency sinusoid and its harmonic, usually with variable
amplitudes, which we interpret as rotational modulation.   A large fraction
(28\,percent) of B stars might be classified as ellipsoidal variables, but
a statistical argument suggests that these are probably rotational 
variables as well.   About 8\,percent of the rotational variables have a 
peculiar periodogram feature which is common among A stars.  The physical 
cause of this is very likely related to rotation.  The presence of so many
rotating variables indicates the presence of star spots.  This suggests that
magnetic fields are indeed generated in radiative stellar envelopes.  
We find five $\beta$~Cep variables, all of which have low frequencies with 
relatively large amplitudes. The presence of these frequencies is a puzzle.  
About half the stars with high frequencies are cooler than the red edge of the 
$\beta$~Cep instability strip.  These stars do not fit into the general 
definition of $\beta$~Cep or SPB variables.  We have therefore assumed they 
are further examples of the anomalous pulsating stars which in the past have 
been called ``Maia'' variables.  We also examined 300 B stars observed in the 
K2 Campaign 0 field.  We find 11 $\beta$~Cep/Maia candidates and many SPB 
variables.  For the stars where the effective temperature can be measured, we 
find at least two further examples of Maia variables. 
\end{abstract}


\section{Introduction}

The {\it Kepler} mission has been an outstanding success, not only in the
detection of extra-solar planets by the transit method, but also in the
general field of astrophysics and asteroseismology.  The superb photometric
precision and almost continuous 4-yr coverage of over 100000 stars has led
to many surprises.  The center of the {\it Kepler} field is about $13^\circ$
above the galactic plane.  B-type main sequence stars are to be found very
close to the galactic plane and, as a result, only a few of the nearest,
brightest, main sequence B stars are visible in the {\it Kepler} field.  There
are a few hundred stars which have effective temperatures $T_{\rm eff} >
10000$\,K according to the {\it Kepler Input Catalogue} (KIC,
\citealt{Brown2011a}).  Most of these have high proper motions and probably are
faint, hot subdwarfs.  Unfortunately, the multicolour photometry used to
derive the stellar parameters in the KIC does not include the U band.  As a
result, the effective temperatures for B stars cannot be trusted.  Therefore
some B stars may have been assigned low temperatures in the KIC, placing them 
among the A stars. 

\citet{Balona2011b} presented the first major study of {\it Kepler} main
sequence B stars based on the first year of operation.  Among the 48 B stars, 
15 stars show the characteristics of SPB variables,  7 of which could be 
considered as hybrid $\beta$~Cep/SPB variables.  The data indicate that
non-pulsating stars exist in the $\beta$~Cep and SPB instability strips.
Apart from the pulsating stars, several different classes of variables can
be identified based on frequency groupings, but the nature of the
variability is unclear.  Subsequently, \citet{McNamara2012} analyzed the
{\it Kepler} light curves of B star candidates to further characterize B-star 
variability.  They found 10 $\beta$~Cep variables and 54 SPB stars among the
sample.  However, many of the $\beta$~Cep candidates are now know from
spectroscopic observations to be hot compact objects and not main sequence B
stars.

The loss of a second spacecraft reaction wheel effectively ended data 
collection in the original {\it Kepler} field after four years of continuous 
monitoring. By pointing near the ecliptic plane, the {\it Kepler} spacecraft 
is able to minimize pointing drift, allowing data collection to proceed with 
much reduced photometric precision.  This mode of operation is called the K2
project.  A preliminary test of this mode for science purpose, Campaign 0, was
made during 2014 Mar 8--27.  The K2 Campaign 0 field, which we will call the
K2-0 field, is closer to the galactic plane ($b = 6^\circ$) and contains many 
more B stars.  However, ground based photometry is very limited and it is 
not possible to estimate effective temperatures for most of the stars.

In this paper we examine in more detail the light curves of B stars observed
in the original {\it Kepler} field using all available (Q0--Q17) data.  We
also include 300 B stars observed in the K2-0 field.  Our aim is to detect 
pulsating main sequence B stars in these two fields by examining the light 
curves and periodograms and to classify them according to variability type
as best as we can.  

It turns out that a large fraction of B stars appear to be rotational 
variables.  This is surprising because B star atmospheres are radiative
which are thought not to be capable of generating the magnetic field
necessary to produce a star spot.  We also find an extraordinary number of 
supposedly ellipsoidal variables. Many of these may be rotational variables as
well.  Furthermore, we show that the strange periodogram feature, first found 
among A stars and described by \citet{Balona2013c}, is also present among the 
B stars.  Finally, we detect several B stars containing high frequencies but 
which are too cool to be classified as $\beta$~Cep variables.  In 
citep{Balona2011b} it was speculated that these anomalous pulsating stars may
be binaries in which the primary is a B star and the secondary a $\delta$~Sct 
variable.  However, the sheer number of such anomalous variables as well as 
recent {\it CoRoT} and ground-based observations do not support this
conclusion.  Because these stars cannot be considered as $\beta$~Cep or SPB
stars and are too hot to be $\delta$~Sct variables, we decided to call them
Maia variables.  This name has been used for such pulsating variables in the
past, but no definite proof of their existence has been found.  Although
they may eventually be explained by known mechanisms, we feel that until
this is done, it is important to distinguish them from the $\beta$~Cep and 
SPB variables.

\begin{figure}
\centering
\includegraphics{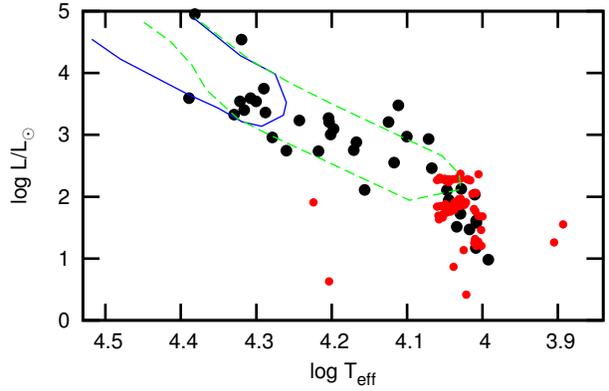}
\caption{Location of stars in the theoretical H-R diagram.  The large
(black) dots are stars with stellar parameters derived from spectroscopic or 
Str\"{o}mgren observations, while the (red) smaller dots are stars with 
parameters from the KIC.  The region enclosed by the solid line is the 
theoretical instability strip for $\beta$~Cep variables.  The dashed region 
is the theoretical instability strip for SPB variables.  These are from 
\citet{Miglio2007} for $Z=0.02$, $l \le 3$ and OP opacities.}
\label{hrdiag}
\end{figure}

\begin{figure*}
\centering
\includegraphics{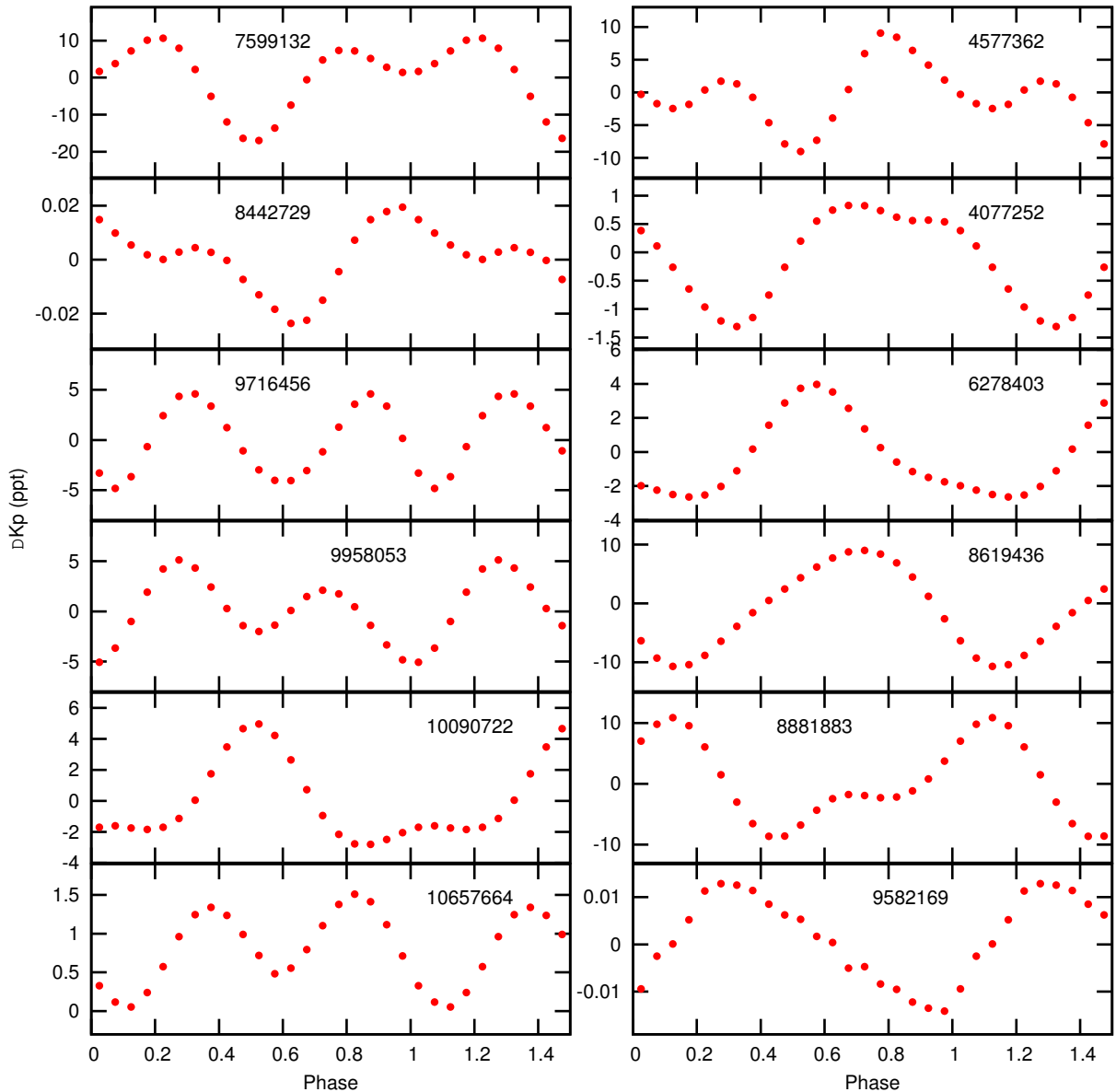}
\caption{Phased light curves of some ellipsoidal variables with
non-sinusoidal variations.}
\label{ell}
\end{figure*}

\begin{figure}
\centering
\includegraphics{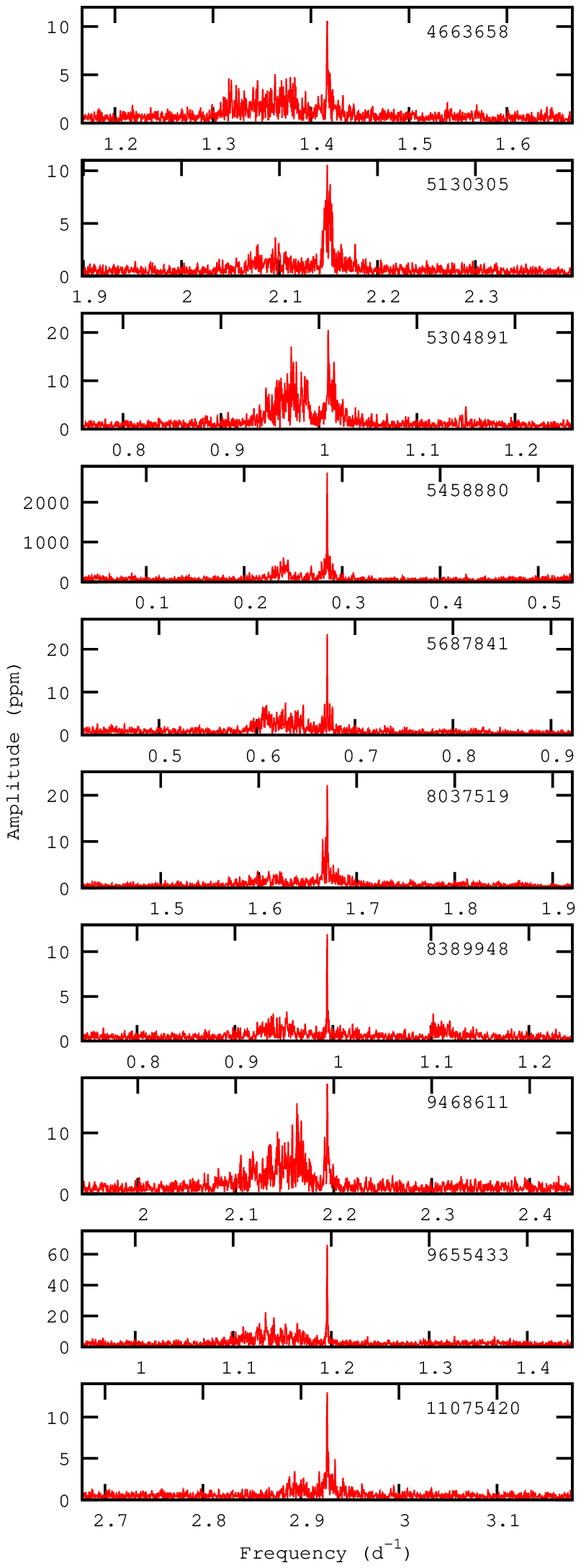}
\caption{Periodograms in the region of the broad and narrow peaks for stars
showing the pattern described by \citet{Balona2014b}.}
\label{mess}
\end{figure}

\section{The data}

The {\it Kepler} light curves are available as uncorrected simple aperture 
photometry (SAP) and with pre-search data conditioning (PDC) in which
instrumental effects are removed \citep{Stumpe2012, Smith2012}.  The vast 
majority of the stars are observed in long-cadence (LC) mode with exposure 
times of about 30\,min. Several thousand stars were also observed in 
short-cadence (SC) mode with exposure times of about 1\,min.  These data 
are publicly available on the Barbara A. Mikulski Archive for Space 
Telescopes (MAST, {\tt archive.stsci.edu}).  

In selecting our sample of known and possible B stars observed by {\it Kepler}
we first  included all stars with effective temperatures $T_{\rm eff} \ge 
10000$\,K in the {\it Kepler} Input Catalogue (KIC, \citealt{Brown2011a}).  
We also added stars in Table\,1 of \citet{Balona2011b} which do not have
effective temperature measurements in the KIC but which are known B stars.
Finally, we performed a literature search on all stars in this list and 
removed compact objects.  The list of stars which we consider to be mostly 
B-type main sequence objects is shown in Table\,\ref{bstars}.  Since the KIC 
effective temperatures are not reliable for B stars, some stars in the table 
may be cooler than B9 and a few may still be unrecognized compact objects. 

In this table, spectroscopic measurements of $T_{\rm eff}$, and surface 
gravity, $\log(g)$, are used whenever possible.  Failing this, values derived 
from Str\"{o}gren photometry are listed.  These parameters are from 
\citet{Balona2011b} which are mostly derived from \citet{Catanzaro2010} 
and \citet{Lehmann2011}.  Further spectroscopic data by \citet{McNamara2012} 
and \citet{Tkachenko2013a, Tkachenko2013b} were also used to update the 
effective temperatures and surface gravities.   When spectroscopic or 
Str\"{o}mgren values are not available, the KIC values are shown marked by a 
colon.   The luminosity $\log(L/L_\odot)$, is calculated from $T_{\rm eff}$ and 
$\log g$ using the relationship in \citet{Torres2010a}.  For those stars where 
only the KIC values are available, the luminosity was derived from the KIC 
effective temperature and radius.  We emphasize, once again, that the KIC 
values are highly unreliable for B stars and they serve merely as indicators 
that the stars may be of type B and nothing more.  Further observations are 
required to ascertain their effective temperatures and surface gravities.  In 
Fig.\,\ref{hrdiag} we plot all the stars of Table\,\ref{bstars} in the 
theoretical H-R diagram.

\section{Classification of variability type}

Classification of variability type is an essential first step in this study.  
We only have the light curve and periodogram, but in most cases this is 
sufficient to enable a variability type to be assigned.  We discuss
the types of variable star that might be found among the B stars.  

Eclipsing binaries are easily classified into two main groups: detached 
systems (Algol type, EA) or semi-detached ($\beta$~Lyr systems, EB).  
In EA systems the components are spherical or slightly ellipsoidal and it is
easy to specify the beginning and end of the eclipses.  Between eclipses the 
light remains almost constant.  In EB systems the stars are tidally distorted 
and it is impossible to specify the exact times of onset and end of eclipses 
because of a continuous change of the combined brightness between eclipses. 

The ellipsoidal variables (ELL) are close binary systems with tidally
distorted components.  Due to the low inclination of the orbital axis, 
there are no eclipses, but the changing aspect towards the observer
results in light variations.  The light variations are a combination of 
tidal distortion, reflection and beaming.  Beaming is induced by the 
stellar radial motion which results in an increase (decrease) in 
brightness when the star is approaching (receding from) the observer.  
The period of the reflection and beaming contributions is the same as 
the orbital period whereas the ellipsoidal effect has half the orbital 
period.

There is also the possibility of rotational modulation in single stars (ROT).
All stars rotate, so it is natural to expect light variations if the
photospheric surface brightness is not perfectly uniform in longitude.  
Several types of Bp stars are known to have photospheric spots, though these 
are supposedly surface abundance variations and not comparable to sun spots.  
We take the view that spots of any description must be a permissible option, 
even though we may not know how they may be formed.  The extraordinary 
photometric precision attained by {\it Kepler} allows detection of very small 
surface brightness variations which may have escaped ground-based observations.
We should also allow for the possibility of differential rotation which 
could result in multiple closely-spaced frequency peaks in the periodogram.

There is a degeneracy between the ELL and ROT classes because it is not
possible to distinguish between orbital and rotational variations without
additional information from spectroscopy.  We  assume that if the 
amplitude of the periodic variation changes in a relatively short time scale
(perhaps a few months), then it is perhaps more likely to be a result of
a star spot.  Orbital variations typically have a longer time scale.  Of 
course, it is equally true that long-lived spots may mimic the light curve
of the ELL class.

Among the sample of stars there are a number of chemically peculiar stars. 
These stars have spots of some kind (perhaps of different abundances) and
are known as $\alpha$~CVn variables (ACV).

We know of two types of pulsating main sequence B stars: the $\beta$~Cep 
(BCEP) variables (high pulsation frequencies) and the slowly pulsating
B stars (SPB stars with low frequencies).  We selected the dividing line 
between low and high frequencies to be around 5\,d$^{-1}$ because very few
known $\beta$~Cep pulsate with lower frequencies and few SPB stars pulsate 
with higher frequencies.  

Some stars break this rule: there are stars with high frequencies which are
cooler than the red edge of the BCEP variables.  It is possible that such
stars are binaries in which the pulsating component is the cooler companion 
of a B star primary.  In this way one could explain these anomalous
pulsating stars as an A-type $\delta$~Sct star orbiting a more luminous
non-pulsating B star \citep{Balona2011b}.  However, it is difficult to
sustain this explanation in the face of the large number of such variables.
In this paper we have provisionally classified such anomalous pulsating 
stars as MAIA variables, a term which is discussed below.  We classify a
star as a MAIA variable if it is a B star with high frequencies that is
cooler than the coolest known $\beta$~Cep variable.  This does not
necessarily imply that Maia variables require a separate driving mechanism.
We use the term because these stars violate the definitions of existing
classes.

It is not always possible to distinguish between the SPB and ROT classes.
It would be incorrect to dismiss the presence of rotational modulation in
any star so that multiple close low frequencies might be a sign of 
differential rotation and need not automatically imply pulsation.  We 
generally assume that pulsation is responsible unless we see at least 
one harmonic, in which case rotation is to be preferred.  The reason is that
harmonics of pulsation modes are not normally visible unless the amplitude
is large, whereas harmonics are a natural consequence of localized spots.

With the above considerations in mind, we assigned a variability type to
each star as best as we could.  Results for the B stars in the {\it Kepler}
field are shown in Table\,\ref{bstars}.  Nearly all the stars in
\citet{Balona2011b} are listed in this table.  However, KIC\,8087269, 
8161798, 8488717 and 12207099 are blue horizontal branch stars 
(R.H. {\O}stensen, priv. comm.), while KIC\,10536147 is a pulsating white 
dwarf \citep{McNamara2012}.  These stars are omitted.

\begin{center}
\begin{table*}
\caption{Candidate main sequence B stars in the {\it Kepler} field.  The
KIC number and a variable star classification is shown in the first two
columns.  Stars showing little signs of variability are indicated by a dash
in the type column.  The Kp magnitude, effective temperature, $T_{\rm eff}$ 
and surface gravity, $\log(g)$, are from the KIC if followed by a colon, 
otherwise it is from \citet{Balona2011b}. The luminosity $\log\tfrac{L}{L_\odot}$,
photometric rotation period and the projected rotational velocity,
$v\sin i$, are also shown.  Remarks are given in the last column.}
\label{bstars}
\begin{tabular}{rlrllllrl}
\hline
\multicolumn{1}{c}{KIC} & 
\multicolumn{1}{c}{Type} & 
\multicolumn{1}{c}{Kp} & 
\multicolumn{1}{c}{$T_{\rm eff}$} & 
\multicolumn{1}{c}{$\log\tfrac{L}{L_\odot}$} & 
\multicolumn{1}{c}{$\log (g)$} & 
\multicolumn{1}{c}{$P_{\rm rot}$} & 
\multicolumn{1}{c}{$v\sin i$} & 
\multicolumn{1}{c}{Notes} \\
\multicolumn{1}{c}{} & 
\multicolumn{1}{c}{} & 
\multicolumn{1}{c}{mag} & 
\multicolumn{1}{c}{K} & 
\multicolumn{1}{c}{} & 
\multicolumn{1}{c}{dex} & 
\multicolumn{1}{c}{d} & 
\multicolumn{1}{c}{km\,s$^{-1}$} & 
\multicolumn{1}{c}{} \\
\hline
   1028085 & ELL         &  12.649 &  10721: &   3.64: &    2.30: & 0.5456 &         &                    \\
   1430353 & SPB/ROT     &  12.391 &  10765: &   3.65: &    2.30: & 1.0179 &         &                    \\
   2987640 & MAIA        &  12.641 &  10985: &   4.13: &    1.81: &        &         &                    \\
   3216449 & ELL         &  11.821 &  10232: &   4.40: &    1.32: & 0.9436 &         &                    \\
   3240411 & SPB/BCEP    &  10.241 &  20980  &   4.01~ &    3.54~ &        &    42.6 & B2\,V                \\
   3343239 & MAIA        &  14.420 &  10691: &   3.93: &    1.97: &        &         &                    \\
   3442422 & ROT         &  11.954 &  10732: &   4.02: &    1.88: & 14.286 &         &                    \\
   3459297 & MAIA        &  12.554 &  10592: &   4.63: &    1.14: &        &         &                    \\
   3629496 & ROT         &   8.334 &  11320  &   3.75~ &    2.40~ &        &   160.0 & B8.5\,IV             \\
   3756031 & SPB/MAIA    &  10.042 &  15980  &   3.75~ &    3.21~ &        &    30.8 & B5\,V/IV             \\
   3839930 & SPB         &  10.864 &  16500  &   4.20~ &    2.73~ &        &         &                    \\
   3848385 & ROT         &   8.958 &  11680  &   3.76~ &    2.46~ & 0.9709 &         & B8\,V                \\
   3865742 & SPB/ROT     &  11.130 &  19500  &   3.70~ &    3.75~ & 4.1118 &   133.0 &                    \\
   4064365 & ELL         &  12.296 &  11281: &   4.17: &    1.84: & 8.9606 &         &                    \\
   4077252 & ELL         &  12.314 &  10514: &   3.95: &    1.91: & 3.9754 &         & KOI\,5038.01        \\
   4276892 & ELL/ROT     &   9.185 &  10800  &   4.10~ &    1.86~ & 6.4309 &    10.0 & B8/A0              \\
   4366757 & ROT         &  11.836 &  11423: &   4.37: &    1.63: & 3.1696 &         &                    \\
   4373805 & ROT         &  13.543 &  11087: &   4.12: &    1.84: & 7.1429 &         &                    \\
   4476114 & SPB/ROT     &  12.921 &  10988: &   4.14: &    1.80: & 2.9494 &         &                    \\
   4577362 & ELL         &  15.514 &  11139: &   3.76: &    2.27: & 1.0949 &         &                    \\
   4581434 & ELL         &   9.163 &  10200  &   4.20~ &    1.61~ & 0.5028 &   200.0 & B9/A5              \\
   4660027 & ELL         &  10.632 &  10121: &   3.46: &    2.36: & 1.5152 &         &                    \\
   4663658 & ROT-d       &  11.321 &  10122: &   4.06: &    1.69: & 0.7059 &         &                    \\
   4931738 & SPB         &  11.645 &  10136: &   4.42: &    1.27: &        &         & Equal spacing      \\
   4936089 & SPB         &  11.928 &  11295: &   4.28: &    1.71: &        &         &                    \\
   4939281 & SPB/MAIA    &  12.075 &  11298: &   4.32: &    1.66: &        &         &                    \\
   5130305 & ROT-d       &  10.202 &  10670  &   3.86~ &    2.13~ & 0.4654 &   155.0 & B9\,IV/V             \\
   5217845 & ELL         &   9.429 &  11790  &   3.41~ &    2.93~ & 0.8391 &   237.0 & B8.5\,III            \\
   5304891 & ROT-d       &   9.172 &  13100  &   3.90~ &    2.55~ & 0.9907 &   180.0 & B4-B8              \\
   5450881 & ROT         &  12.468 &  10197: &   4.43: &    1.27: & 0.2940 &         & Equal spacing      \\
   5458880 & ROT-d       &   7.824 &  24070  &   3.18~ &    4.95~ & 3.5112 &         & B0\,III              \\
   5477601 & ROT         &  12.793 &  10906: &   4.14: &    1.78: & 5.2002 &         &                    \\
   5479821 & ROT         &   9.886 &  14810  &   3.97~ &    2.75~ & 1.7012 &    85.0 & B5.5\,V, He-weak     \\
   5530935 & ELL         &   8.040 &  10201: &   3.75: &    2.06: & 0.7852 &         & B9\,V                \\
   5557097 & -           &  14.545 &  10454: &   3.60: &    2.28: &        &         &                    \\
   5687841 & ROT-d       &  10.385 &  10708: &   3.64: &    2.30: & 1.4885 &         &                    \\
   5706079 & ROT         &  11.581 &  10680: &   4.03: &    1.85: & 4.2550 &         &                    \\
   5786771 & ROT         &   9.139 &  10700  &   4.20~ &    1.72~ & 0.4041 &   200.0 & B9/A5              \\
   5941844 & SPB         &   9.283 &  11443: &   4.32: &    1.69: &        &         & B9                 \\
   6065699 & ELL         &   7.833 &  10900: &   4.12: &    1.80: & 3.9872 &         & B9                 \\
   6128830 & ROT         &   9.260 &  12600  &   3.50~ &    2.97~ & 4.8426 &    15.0 & B6\,HgMn            \\
   6205507 & -           &  12.692 &  10925: &   4.92: &    0.87: &        &         &                    \\
   6214434 & ROT         &  13.049 &  11099: &   4.17: &    1.79: & 3.0    &         &                    \\
   6278403 & ELL         &   8.758 &  10687: &   4.01: &    1.88: & 1.1911 &         & B9                 \\
   6780397 & ROT         &  10.051 &  11166: &   4.17: &    1.81: & 0.8529 &         & B9                 \\
   6848529 & ROT         &  10.734 &  19300  &   3.80~ &    3.60  &        &    10.0 &                    \\
   6950556 & ELL         &  12.751 &  10942: &   3.70: &    2.28: & 1.5116 &         &                    \\
   6954726 & BE          &  11.930 &         &         &          & 0.9700 &   160.0 & StHA166, Be         \\
   7599132 & ELL         &   9.394 &  11090  &   4.08~ &    1.95~ & 1.3036 &    63.0 & B8.5\,V              \\
   7749504 & ROT         &  12.723 &  11064: &   3.73: &    2.28: & 0.2700 &         &                    \\
   7778838 & ELL         &  11.868 &  11030: &   4.18: &    1.76: & 5.8754 &         &                    \\
   7974841 & ELL/ROT     &   8.198 &  10650  &   3.87~ &    2.11~ & 2.8433 &    33.0 & B8\,V, B7\,IV/V        \\
   8018827 & EB          &   8.065 &  10945  &   3.98~ &    2.04~ & 0.3979 &   243.0 & B9, B8.5\,IV/V       \\
   8037519 & ROT-d       &   8.006 &  11089: &   3.78: &    2.23: & 0.5988 &         & B8\,V                \\
   8057661 & BE/BCEP     &  11.609 &  21360  &   4.23~ &    3.33~ &        &    49.0 & Be                 \\
   8129619 & ELL         &  10.992 &  10203: &   4.49: &    1.20: & 0.9519 &         &                    \\
   8167938 & SPB         &  10.846 &  11167: &   4.23: &    1.74: &        &         & KOI\,6052.01        \\
\hline
\end{tabular}
\end{table*}
\end{center}

\setcounter{table}{0}

\begin{center}
\begin{table*}
\caption{Continued.}
\begin{tabular}{rlrllllrl}
\hline
\multicolumn{1}{c}{KIC} & 
\multicolumn{1}{c}{Type} & 
\multicolumn{1}{c}{Kp} & 
\multicolumn{1}{c}{$T_{\rm eff}$} & 
\multicolumn{1}{c}{$\log\tfrac{L}{L_\odot}$} & 
\multicolumn{1}{c}{$\log (g)$} & 
\multicolumn{1}{c}{$P_{\rm rot}$} & 
\multicolumn{1}{c}{$v\sin i$} & 
\multicolumn{1}{c}{Notes} \\
\multicolumn{1}{c}{} & 
\multicolumn{1}{c}{} & 
\multicolumn{1}{c}{mag} & 
\multicolumn{1}{c}{K} & 
\multicolumn{1}{c}{} & 
\multicolumn{1}{c}{dex} & 
\multicolumn{1}{c}{d} & 
\multicolumn{1}{c}{km\,s$^{-1}$} & 
\multicolumn{1}{c}{} \\
\hline
   8177087 & SPB         &   8.093 &  13330  &   3.42~ &    3.21~ &        &    22.2 & B7\,III He-strong    \\
   8183197 & ELL         &  11.223 &  10822: &   4.07: &    1.84: & 3.6914 &         &                    \\
   8264293 & ROT         &  11.310 &  10038: &   4.46: &    1.21: & 0.3400 &         &                    \\
   8324268 & ACV         &   7.955 &  11370  &   3.35~ &    2.93~ & 2.0090 &    31.0 & B9pSiCr, V2095\,Cyg \\
   8351193 & ROT         &   7.598 &  ~9980  &   3.80~ &    2.05~ & 0.5685 &   180.0 & B9\,V Si             \\
   8381949 & SPB/BCEP    &  11.104 &  24500  &   4.30~ &    3.59~ &        &         &                    \\
   8389948 & ROT-d       &   9.220 &  10240  &   3.86~ &    2.03~ & 1.0060 &   142.0 & B9.5\,V/IV           \\
   8442729 & ELL         &  10.905 &  10753: &   3.95: &    1.96: & 1.6576 &         &                    \\
   8459899 & SPB         &   8.730 &  15760  &   3.81~ &    3.10~ &        &    53.0 & B4.5\,IV, SB2?       \\
   8559392 & ROT         &  12.983 &  10510: &   5.23: &    0.42: & 3.0000 &         &                    \\
   8619436 & ELL         &  11.806 &  10694: &   3.64: &    2.29: & 3.1328 &         &                    \\
   8692626 & ELL         &   8.342 &  ~9826  &   4.71~ &    0.98~ & 1.6483 &         & B9                 \\
   8714886 & SPB/MAIA    &  10.945 &  19000  &   4.30~ &    2.96~ &        &         &                    \\
   8766405 & ROT         &   8.875 &  12930  &   3.16~ &    3.48~ & 0.5457 &   240.0 & B7\,III              \\
   8881883 & ELL         &  12.441 &  10710: &   3.63: &    2.31: & 3.3234 &         &                    \\
   9005047 & ROT         &  11.686 &  10477: &   3.61: &    2.28: & 1.3167 &         &                    \\
   9227988 & SPB         &  12.845 &  10890: &   4.02: &    1.91: &        &         &                    \\
   9278405 & ROT         &  10.243 &  11486: &   4.20: &    1.84: & 0.5590 &         & B9                 \\
   9468611 & ROT-d       &  13.144 &  11063: &   3.73: &    2.27: & 0.4559 &         &                    \\
   9582169 & ELL         &  12.051 &  11373: &   3.79: &    2.28: & 0.4006 &         &                    \\
   9655433 & MAIA/ROT-d  &  11.930 &         &         &          & 0.8361 &         & NGC\,6811, B7/A3     \\
   9715425 & SPB         &  13.065 &  11199: &   4.11: &    1.88: &        &         &                    \\
   9716301 & ROT         &  11.622 &  ~7822: &   3.63: &    1.55: & 1.9940 &         & NGC\,6811, B9/A3     \\
   9716456 & ELL         &  11.980 &  ~8035: &   3.94: &    1.26: & 1.8132 &         & A1/B9              \\
   9725496 & -           &  12.256 &  10257: &   4.46: &    1.25: &        &         &                    \\
   9884329 & ELL         &  12.144 &  10695: &   3.57: &    2.37: & 0.3692 &         &                    \\
   9958053 & ELL         &  11.727 &  10225: &   4.01: &    1.77: & 2.7334 &         &                    \\
   9964614 & SPB/BCEP    &  10.806 &  20300  &   3.90~ &    3.59~ &        &         &                    \\
  10090722 & ELL         &  12.998 &  11383: &   3.78: &    2.29: & 6.0100 &         &                    \\
  10118750 & SPB/ROT     &  13.897 &  11147: &   3.75: &    2.28: & 0.2700 &         &                    \\
  10130954 & ELL         &  11.134 &  19400  &   4.00~ &    3.36~ & 2.2467 &         &                    \\
  10220209 & SPB         &  14.120 &  15969: &   6.10: &    0.63: &        &         &                    \\
  10285114 & SPB/MAIA    &  11.232 &  18200  &   4.40~ &    2.74~ &        &         &                    \\
  10415184 & -           &  14.464 &  10265: &   3.99: &    1.80: &        &         &                    \\
  10526294 & SPB         &  13.033 &  11072: &   3.74: &    2.27: &        &         &                    \\
  10657664 & ELL         &  13.234 &  10564: &   3.99: &    1.87: & 3.2733 &         & KOI\,964.01        \\
  10658302 & SPB         &  13.125 &  15900  &   3.90~ &    3.01~ &        &         & Equal spacing      \\
  10671396 & -           &  10.479 &  11030: &   4.12: &    1.84: &        &         &                    \\
  10684843 & -           &  10.472 &  10000: &   4.04: &    1.68: &        &         &                    \\
  10790075 & ROT?        &  12.808 &  11396: &   3.77: &    2.31~ & 3.9060 &         &                    \\
  10797526 & SPB         &   8.342 &  20873  &   3.22~ &    4.54~ &        &         & B5/A5              \\
  10960750 & SPB/BCEP    &   9.844 &  19960  &   3.91~ &    3.54~ &        &   253.0 & B2.5\,V              \\
  10965032 & SPB/ROT     &  14.647 &  11488: &   3.82: &    2.27: & 0.3000 &         &                    \\
  11038518 & ELL         &  13.554 &  11224: &   3.75: &    2.29: & 0.7646 &         &                    \\
  11075420 & ROT-d       &  11.604 &  11089: &   3.74: &    2.28: & 0.3417 &         &                    \\
  11360704 & SPB/ROT     &  10.738 &  20700  &   4.10~ &    3.40~ & 0.4060 &         &                    \\
  11454304 & SPB/MAIA    &  12.951 &  17500  &   3.90~ &    3.23~ &        &         &                    \\
  11565279 & -           &  12.324 &  11374: &   4.17: &    1.85: &        &         &                    \\
  11671923 & SPB         &  10.679 &  10044: &   4.24: &    1.46: &        &         &                    \\
  11817929 & -           &  10.381 &  16000  &   3.70~ &    3.27~ &        &         & A0                 \\
  11912716 & SPB/ROT     &   6.602 &  10386: &   3.61: &    2.26: & 1.5106 &         & B9\,III              \\
  11957098 & ELL         &  12.887 &  11183: &   3.78: &    2.26: & 1.7367 &         &                    \\
  11973705 & MAIA        &   9.116 &  11150  &   3.96~ &    2.11~ &        &   103.0 & B8.5\,V/IV, SB2      \\
  12060050 & ROT         &  13.671 &  10825: &   3.67: &    2.29: & 3.0300 &         &                    \\
  12217324 & ROT         &   8.312 &  10380  &   3.75~ &    2.20~ & 6.6600 &    19.0 & B9.5\,IV             \\
  12258330 & SPB/MAIA    &   9.517 &  14700  &   3.85~ &    2.88~ &        &   130.0 & B5\,V           \\
  12268319 & ELL         &  11.628 &  10763: &   4.03: &    1.87: & 3.6311 &         &                    \\
  12453485 & ROT         &  12.361 &  10317: &   3.79: &    2.05: & 0.4135 &         &                    \\
\hline                                                       
\end{tabular}                                                           
\end{table*}
\end{center}

\section{Eclipsing binaries}

There are surprisingly few eclipsing binaries - only one in our sample of
115 stars.  KIC\,8018827 shows the characteristic light curve of a contact 
binary with a secondary eclipse that is approximately one-quarter the depth 
of primary eclipse.  The orbital period $P = 0.3979$\,d is very short 
\citep{Tkachenko2013b}.  If the stellar mass is approximately 3\,$M_\odot$
the semi-major axis will be approximately 3.3\,$R_\odot$.  It is classified 
as a rotational variable in \citet{Balona2011b}.

\section{Ellipsoidal variables}

The periodogram of an ellipsoidal binary consists of sharp peaks at the
fundamental frequency and its harmonics.  Usually only the first harmonic is
visible.  Sometimes the fundamental has a lower amplitude than the first 
harmonic.  As a consequence of differences in harmonic content, the shapes of 
the light curves can be very different.  About half the ellipsoidal variables 
in Table\,\ref{bstars} have almost sinusoidal light curves.  Examples of ELL
variables with non-sinusoidal light curves are shown in Fig.\,\ref{ell}.
These were obtained by phasing the light curve with the fundamental period and
averaging the phase curve in twenty phase bins.  The large number of ELL
variables is rather surprising because one expects rather few systems at low 
inclinations.  The highly non-sinusoidal light curves in Fig.\,\ref{ell}
may be a result of a strong reflection effect and/or beaming.  A study of the 
stellar parameters required to produce such light curves is beyond the scope 
of this paper, but is clearly of great interest given the considerable number 
of these systems among the {\it Kepler} B stars.

In KIC\,3216449 the amplitude increases gradually from about 2.5\,ppt (parts
per thousand) to 3.5\,ppt over the 4-yr observing period.  In  KIC\,4077252 
the amplitude decreases from 12 to 9\,ppt over the 4-yr period.   In KIC\,4064365 
there is a single strong peak at 0.1116\,d$^{-1}$ and another much weaker one 
at 0.0881\,d$^{-1}$.  The secondary peak could indicate non-synchronous 
rotation.  In KIC\,4276892 the dominant peak is $\nu = 0.1555$\,d$^{-1}$ with 
a much weaker peak at 0.0439\,d$^{-1}$.  Both peaks are fairly broad and the 
light curve is rather complex.  It is difficult to understand the variation 
in terms of a simple ellipsoidal system and perhaps this star should be
classified as a rotational variable.

The stars classified as ELL in Table\,\ref{bstars} are nearly always
classified as probable binaries in \citet{Balona2011b}.  With the much larger 
data set used here, one is able to determine the orbital period and shape of 
the light curve quite accurately.  For KIC\,7974841 \citet{Balona2011b} adopt 
a classification of ROT/SPB.  The periodogram shows a sharp peak ($P = 
2.8433$\,d) and harmonic, but also a broad peak at $P = 4.0$\,d and its 
harmonic.  This does not look like the frequency pattern of an SPB star 
but could be a spotted companion in a non-synchronous orbit.  
\citet{Tkachenko2013b} could not find evidence for binarity from their 
two spectra.  We give it the classification ELL/ROT.

\section{Rotational variables}

A characteristic of the ROT type is a variation in amplitude on a timescale
of months.  Often there are beating effects and traveling features in the 
light curve which is a characteristic feature of late-type variables with 
star spots.  An interesting subset of the rotational variables consists of a 
pattern first described in \citet{Balona2014b}.  Examples of this pattern are 
shown in Fig.\,\ref{mess}.  As can be seen, there is a broad peak flanked by a
sharp peak at a slightly higher frequency.  For convenience we classify stars 
with this pattern as ROT-d because the pattern is similar to the letter d. 

\citet{Balona2013c} discovered that about 20\,percent of {\it Kepler} A 
stars are ROT-d variables.  \citet{Balona2014b} suggested that the broad 
feature is due to differential rotation and the sharp peak a result of 
reflection from a planet in a co-rotating orbit (not necessarily transiting).  
Whatever the reason, ROT-d variables are very common and require an
explanation.  Fig.\,\ref{mess} shows the periodograms of all ROT-d stars in
our sample.

In KIC\,9655433 three of four peaks with amplitudes in the range 0.2--1.2\,ppt 
can be seen with $\nu > 22.5$\,d$^{-1}$.  The high frequencies and low
temperature means that this is a MAIA variable.  There are scattered 
low-amplitude peaks all the way to zero frequency.  There is also a ROT-d 
pattern with the sharp peak at 1.1960\,d$^{-1}$ and broad peak at 
$\nu_B = 1.14$\,d$^{-1}$ and a hint of the harmonic.  

In KIC\,4366757 there are two sharp, low-amplitude peaks at  $\nu_1 = 
0.3155$ and $\nu_2 = 2.7905$\,d$^{-1}$.   The lower of the two frequencies 
is  embedded in a broad peak and there is some weak structure at $2\nu_1$ 
suggesting rotational modulation.  In KIC\,4373805 there is just a single, 
broad, very weak peak at 0.140\,d$^{-1}$, though no harmonic is visible.  
This period is clearly visible in the light curve which is characterized 
by variable amplitude and an overall decrease in amplitude from about 
0.9 to 0.6\,ppt over four years.  

In KIC\,4476114 the highest peak is at $\nu_1 = 0.3390$\,d$^{-1}$ with 
smaller peaks at $\nu_2 = 0.1647$ and $\nu_3 = 0.8250$\,d$^{-1}$.  There 
is no obvious relationship between these three frequencies except 
$\nu_3 = 5\nu_2$.  If the term SPB is a catch-all for unexplained peaks, then 
perhaps this could be classified as a peculiar SPB with only three unstable 
frequencies.  However, the light curve is typical of late-type stars with 
spots, so perhaps this is a manifestation of rotational modulation.
The light curve of KIC\,5450881 also looks very much like that of a 
late-type spotted star.  The periodogram is dominated by three peaks 
$\nu_1 = 3.6787$, $\nu_2 = 3.2700$, $\nu_3 = 4.0874$\,d$^{-1}$ and the 
harmonic $2\nu_1$.  

In KIC\,8264293 there is a tight group of equidistant peaks at  $\nu = 
2.9556$, 3.1295, 2.4091, 2.9309, 2.4228, 2.8519\,d$^{-1}$, with equal spacing
($\delta\nu = 0.023$\,d$^{-1}$).  The first harmonics of these peaks are also
visible.  This may be a rotational variable with splitting caused by
the proximity effect of a binary with orbital frequency 0.023\,d$^{-1}$.

In the B7\,III star KIC\,8766405 there are about a dozen peaks below 6\,d$^{-1}$ 
with the highest at $\nu_1 = 1.8326$\,d$^{-1}$, amplitude 1.7\,ppt.  A 
strong harmonic is present for $\nu_2 = 0.8658$, which suggests that this 
may be the rotation frequency.  The peak at $\nu_3 = 1.9774$ is also  strong.  
The harmonics $2\nu_1$ and $2\nu_3$ have relatively large amplitudes.  All 
this is more suggestive of differential rotational modulation than SPB
pulsation, but either (or both) is possible.

\begin{figure}
\centering
\includegraphics{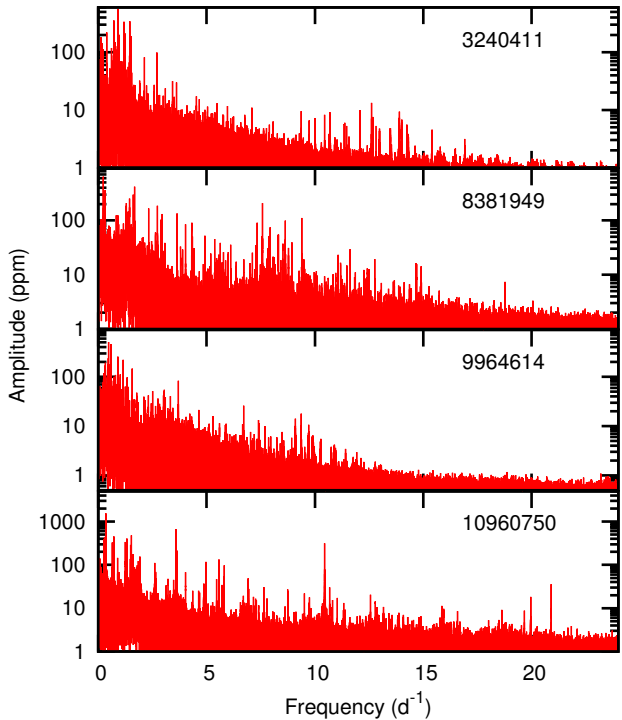}
\caption{Periodograms of stars classified as SPB/$\beta$~Cep hybrids.
A logarithmic amplitude scale is used to show the very weak high frequencies.}
\label{bcep}
\end{figure}

\section{Chemically peculiar stars}

There are four known chemically peculiar stars in our sample and all show
rotational modulation.  The only star of the four which has been given the
designation ACV in the {\it General Catalogue of Variable Stars}
\citep{Samus2009} is KIC\,8324268, but the other three stars should probably 
be given the same classification.  KIC\,8324268 is a B9pSiCr star with a peak 
at $\nu = 0.4977$\,d$^{-1}$ and amplitude 12\,ppt. \citet{Tkachenko2013b} 
find line profile variations consistent with abundance spots. 

\citet{Tkachenko2013b} find that KIC\,8351193 has a metallicity, 
[M/H]$<$ -0.8\,dex, with nearly solar helium abundance and overabundance of 
Mg compared to the derived Fe content.  In addition there is a strong 
enhancement of Si. The light curve shows a very low amplitude (2 ppm) 
rather broad peak at  $\nu = 1.7589$\,d$^{-1}$ and its harmonic which we 
interpret as rotational modulation.

KIC\,5479821 is a B5.5\,V He-weak star \citep{Lehmann2011}. The periodogram 
shows a dominant peak at $\nu_1 = 0.5878$\,d$^{-1}$ and first harmonic.  
Higher harmonics are present but very weak.  In addition, there is another 
peak at $\nu_2 = 0.1841$\,d$^{-1}$ which bears no relationship to $\nu_1$.  
The light curve is sinusoidal with minima and maxima of slightly different 
amplitude superimposed on the low-frequency $\nu_2$ variation.  We 
presume that $\nu_1$ is the rotation frequency.  

KIC\,6128830 is a sharp-lined B6\,HgMn star \citep{Catanzaro2010}.  There is 
only one frequency $\nu = 0.2065$\,d$^{-1}$ and its harmonic.

\section{$\beta$~Cep variables}

Pulsations in $\beta$~Cep and SPB stars are driven by the $\kappa$ opacity
mechanism due to the iron-group opacity bump at about 200\,000\,K.
This mechanism can destabilize p modes in $\beta$~Cep stars and g modes in 
SPB stars.  Driving of pulsations can only occur if certain criteria are met. 
One of these criteria is that the pulsation period is of the same order as 
the thermal timescale, otherwise the driving region remains in thermal 
equilibrium and cannot absorb/release the heat required for driving the 
pulsations. Another requirement is that the pressure variation is large and 
varies only slowly within the driving region.  This criterion is satisfied 
for low radial order p modes in $\beta$~Cep stars.  In the relatively shallow 
partial ionization zone of iron-group elements the thermal timescale is well 
below 1\,d. It turns out the timescale constraint is fulfilled only for p 
modes of low radial and mixed modes as well as for g modes with $l \geq
4$, though it is possible to find unstable g modes with $l$ as low as $l=2$
for models close to the core hydrogen burning phase.  In less massive 
stars, the iron opacity bump is located deeper and, in addition, the luminosity 
is lower. Both factors contribute to an increase in the thermal timescale. 
Consequently, all p modes are stable because their periods are much shorter 
than the thermal timescale, while g modes fulfill the timescale 
and pressure amplitude constraints and are unstable. Thus we find two 
instability regions among the B stars: the $\beta$~Cep instability strip 
and, at lower masses, the SPB instability strip.

In general, g modes are heavily damped in B stars of high mass because they 
have high amplitudes in the deep layers just above the convective core. The 
temperature variation gradient is very large in this region, leading to 
significant heat loss and damping of the pulsations. However, the structure of
the eigenfunction plays an important role and for some modes driving due to the 
iron-group opacity bump exceeds damping in the inner layers.  Hence some 
low-frequency g modes can occur together with high frequency p modes in the 
same star.  Stars where both low-frequency SPB pulsations and high-frequency 
$\beta$~Cep pulsations are present are called SPB/$\beta$~Cep hybrids.

In Table\,\ref{bstars} there are four stars which show high frequencies and
lie within the known $\beta$~Cep instability strip.  The periodograms of
these stars are shown in Fig.\,\ref{bcep}.  In addition, the Be star
KIC\,8057661 discussed below appears to be a $\beta$~Cep variable.  
KIC\,3240411 is a sharp-lined B2\,V star whose effective temperature was 
measured spectroscopically by \citet{Lehmann2011}.  The star has a large 
number of peaks below 2\,d$^{-1}$, although significant very low-amplitude 
peaks are present up to 17\,d$^{-1}$.  The star would have been classified as 
a simple SPB if only peaks with amplitudes greater than 50\,ppm are considered,
but otherwise as a SPB/$\beta$~Cep hybrid.  

KIC\,10960750 is a rapidly-rotating B2.5\,V star \citep{Lehmann2011}. Peaks as 
high as 21\,d$^{-1}$ are visible, though most of the higher amplitudes are in 
the range 0--4\,d$^{-1}$.  KIC\,8381949 shows quite a rich frequency spectrum 
with significant peaks at least as high as 15\,d$^{-1}$ and a moderate 
amplitude group in the interval 7--10\,d$^{-1}$.   

\begin{figure}
\centering
\includegraphics{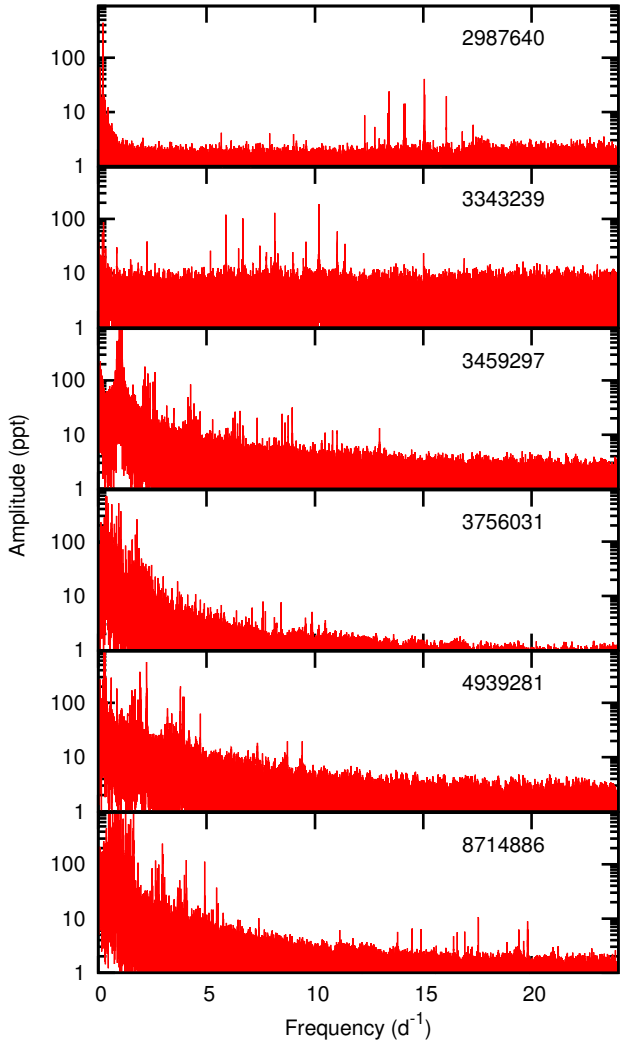}
\caption{Periodograms of stars classified as MAIA variables.
A logarithmic amplitude scale is used to show the very weak high frequencies.}
\label{maia1}
\end{figure}

\section{The Maia variables}

\citet{Struve1955} suggested the possibility of a new class of short-period 
late B or early A variable stars on the basis of radial velocity observations 
of the star Maia, a member of the Pleiades.  Later, \citet{Struve1957} 
disclaimed such variability in Maia, although short-term variations were found
in the strengths of the helium lines.  Since that time, several attempts have 
been made to detect short-term periodic variability in stars lying between the
$\beta$~Cep and $\delta$~Sct instability strips without success 
\citep{McNamara1985, Scholz1998, Lehmann1995}.    Several hundred Maia 
candidates were proposed by \citet{Percy2000} in an analysis of {\it HIPPARCOS}
data.  Some 35 candidates were subsequently observed by \citet{DeCat2007}, but
all were reclassified into other variability classes. 

In spite of these failures, there are persistent reports that such variables
do exist.  A good example is $\alpha$~Dra, an A0\,III Maia candidate with 
a period of about 53\,min and very small amplitude \citep{Kallinger2004}. 
Further examples of possible Maia candidates are provided by {\it CoRot} 
observations of B-type variables \citep{Degroote2009b}. These stars have a 
very broad range of frequencies and low amplitudes.  In addition, several SPB 
stars with residual excess power at frequencies typically a factor three above
their expected g-mode frequencies were found by {\it CoRot}.  The high 
frequencies found in some SPB stars observed by {\it Kepler} \citep{Balona2011b} 
may be taken as further examples of these anomalous variables.  

There is no doubt that cool B stars with high frequencies do exist, although
they have not been given the name ``Maia'' variables. It may be argued that 
perhaps the temperatures of these stars are not correct or perhaps some are 
binary systems in which the fainter component is a $\delta$~Sct star.  It 
would certainly be important to obtain spectroscopic observations of these 
stars to detect the supposed $\delta$~Sct companion.  However, the sheer 
number of such stars now known argues strongly against such an interpretation.

Further evidence indicating the existence of these anomalous pulsating
stars has recently been found by \citet{Mowlavi2013} in photometric 
observations of the young open cluster NGC\,3766.  They found a large 
population (36 stars) of new variable stars between the red edge of the
SPB instability strip and the blue edge of the $\delta$~Sct instability
strip.  Most stars have periods in the range 0.1--0.7\,d, with
amplitudes between 1--4 mmag.  About 20\,percent of stars in this region 
of the H-R diagram were found to be variable.  They suggest that rapid
rotation may be a requirement for the presence of these modes.

Since there is a possibility of some unknown pulsation mechanism operating
in these stars and since high frequencies in late B stars are unexplained
in terms of our current understanding (see below), we decided to classify 
all stars with high frequencies cooler than about 20000\,K as MAIA variables.  
Whether such stars require a separate driving mechanism is, as yet, not
clear.  It may be possible that they could be assimilated into the SPB class
when rapid rotation is taken into account.   Periodograms of some of these 
stars are shown in Fig.\,\ref{maia1}.  

KIC\,3756031 is a sharp-lined B5V/IV star \citep{Lehmann2011}.  Most peaks 
are in the range $\nu < 2$\,d${-1}$, but significant peaks are visible as high 
as  $\nu = 10$\,d$^{-1}$.   In KIC\,11454304 peaks up to 20\,d$^{-1}$ are 
visible, but most peaks are below 5\,d$^{-1}$.  KIC\,12258330 is a He-strong 
star \citep{Lehmann2011}. Peaks up to 14\,d$^{-1}$ are present.

\begin{figure}
\centering
\includegraphics{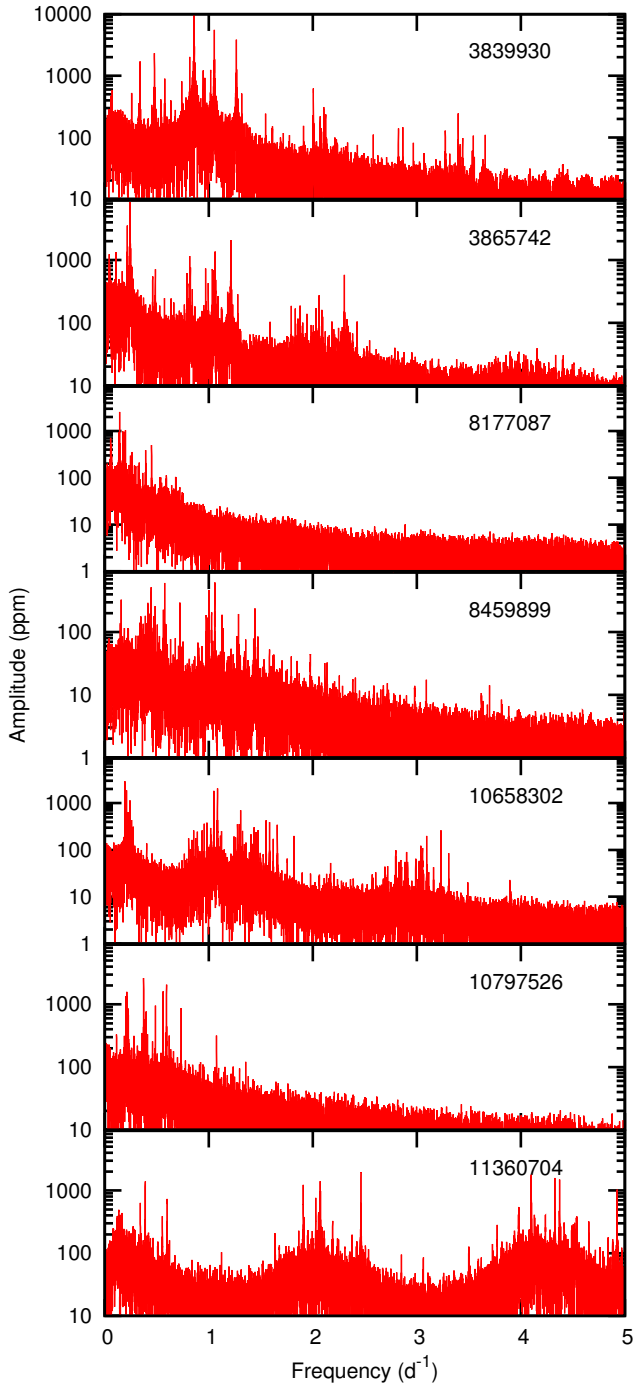}
\caption{Periodograms of some stars classified as SPB variables.}
\label{spb}
\end{figure}

\begin{figure}
\centering
\includegraphics{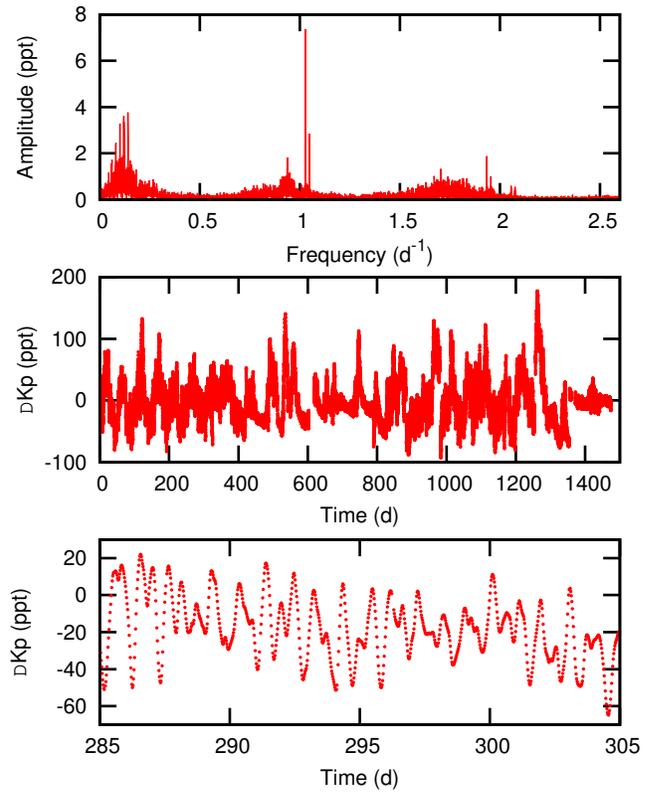}
\caption{Top panel: periodograms of the Be star KIC\,6954726.  Middle panel:
the complete PDC light curve showing moderate outbursts.  Bottom panel: part
of the light curve showing regular variations with variable amplitude.}
\label{006954726}
\end{figure}

\begin{figure}
\centering
\includegraphics{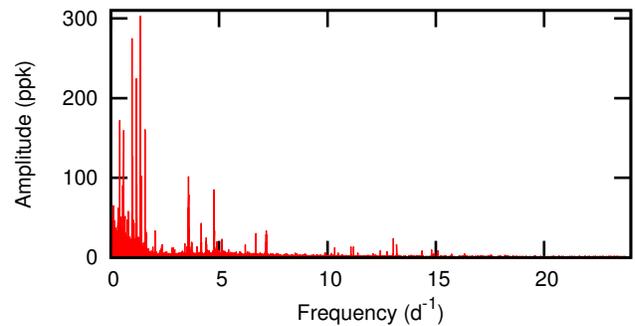}
\caption{Periodogram of the Be star KIC\,8057661.}
\label{008057661}
\end{figure}

\section{SPB variables}

There are 25 stars in Table\,\ref{bstars} which could be considered pure SPB 
variables.  Periodograms of some of these stars are shown in
Figs.\,\ref{spb}. We discuss some of the more interesting stars.

In KIC\,1430353 the peak at $\nu_1 = 0.9824$\,d$^{-1}$ could be the rotational 
frequency since a small peak is present at $2\nu_1$.  The doublet at 
$\nu_2 = 1.8107$, $\nu_3 = 1.82230$ and $\nu_4 = 0.2533$\,d$^{-1}$ as well as 
three weak doublets at 0.144, 0.832, 1.564 with separations 0.0114\,d$^{-1}$ 
(which is the same as $\nu_2 - \nu_3$) have no obvious explanation.  All 
peaks in KIC\,3865742 have $\nu < 2.7$\,d$^{-1}$.  The harmonic of the 
highest peak at $\nu = 0.2432$\,d$^{-1}$ is present, so perhaps one may 
regard this as an SPB/ROT variable.  In \citet{Balona2011b} the star is not
recognized as SPB but as one of those stars belonging to a frequency
grouping.

The periodogram of KIC\,4931738 is quite extraordinary.  There are three main 
peak groups at 0.45, 0.75 and 1.05\,d$^{-1}$.  Each of these groups is comprised 
of a set of several equally-spaced peaks with $\delta\nu = 0.037$\,d$^{-1}$.  
Perhaps this is a proximity effect in a binary where $\delta\nu$ is the orbital 
frequency.  In KIC\,4936089 the periodogram consists of a series of peaks 
in the range $0.8 < \nu < 1.6$\,d$^{-1}$ with amplitudes diminishing with 
increasing frequency.  The light curve shows strong beating with a period of 
about 31\,d.  KIC\,8177087 is a sharp-lined B7\,III He-strong star 
\citep{Lehmann2011}.  There are several peaks with $\nu < 0.46$\,d$^{-1}$.

KIC\,9715425 has a very peculiar light curve showing large excursions or 
flare-like outbursts typically lasting 10\,d and amplitudes of about 
20--80\,ppt.  There is a continuous underlying variation with a period of 
around 1\,d with beating and changes in amplitude.  After JD\,2455500 the 
behaviour changes.  The outbursts stop and the short-period amplitude
becomes larger at about 15\,ppt gradually decreasing to 10\,ppt at the end of 
the observing run.  The periodogram of the whole data string is dominated by a
group of frequencies below 2\,d$^{-1}$.  The largest peak is at 
$\nu_1 =  0.9638$\,d$^{-1}$ with a very weak harmonic.  Harmonics of two 
small peaks at 0.7440 and 0.7747\,d$^{-1}$ can also be seen.  In addition,
a small group of peaks around $\nu_2 = 1.6081$\,d$^{-1}$ is visible.

In KIC\,10118750 the periodogram is dominated by six close peaks with the 
strongest at $\nu_1 = 3.6302$\,d$^{-1}$. The first harmonic of $\nu_1$ is 
strong, suggesting that this may be the rotation frequency.  The periodogram 
of  KIC\,10658302 is remarkable.  There is a group of 6 peaks starting at
$\nu_1 = 0.1931$\,d$^{-1}$ and equally spaced with $\delta\nu = 
0.0161$\,d$^{-1}$.    The remaining peaks all lie in the range 
0.8--1.8\,d$^{-1}$ and show an interesting pattern with the same value 
of $\delta\nu$.  This may be a proximity effect on an SPB binary.

\section{Be stars}

Be stars are B stars in which emission or partial emission is present in
some Balmer lines (strongest in the  H$\alpha$ and the H$\beta$ lines).  The
emission is due to a disk of gas surrounding the star.  The disk is created
through outbursts which occur from time to time.  The reason for the
outbursts is not known.  The most popular view is that the stars are close
to critical rotation and that pulsations are somehow responsible for
triggering the mass loss \citep{Rivinius2013}.  An alternative view is that
the outbursts are a result of stellar activity \citep{Balona2013f}.

In KIC\,6954726 (StHA166) two close, sharp peaks at $\nu_1 = 1.0279$ and 
$\nu_2 =  1.0479$\,d$^{-1}$ are slightly displaced from a broad peak at 
0.94\,d$^{-1}$.  Two other broad features at about 0.12 and 1.75\,d$^{-1}$ 
seem to be symmetrically placed.  The harmonics $2\nu_1, 2\nu_2$ are present 
with small amplitudes.  Fig.\,\ref{006954726} shows the periodogram of the 
whole data set. The full light curve shows what appears to be semi-regular 
outbursts which may be responsible for the broad peak at about 0.12\,d$^{-1}$.
Semi-regular light variations with a period of about 1.0\,d are visible in 
the light curve (bottom panel of the figure).

KIC\,8057661 is another Be star.  Peaks are visible all the way to about 
20\,d$^{-1}$ (Fig.\,\ref{008057661}).  However, the peaks of highest amplitude 
are all below 5\,d$^{-1}$.  The  system of peaks below 1.7\,d$^{-1}$ seem to 
be roughly equally spaced with $\delta\nu \approx 0.17$\,d$^{-1}$.  This may 
be a $\beta$~Cep star which is also a Be star.  Several such Be/$\beta$~Cep 
variables are know of which HD~49330 \citep{Huat2009} is a good example.  
A full analysis of these light curves is beyond the scope of this paper.

\section{Location of $\beta$~Cep and SPB stars in the instability strip}

\begin{figure}
\centering
\includegraphics{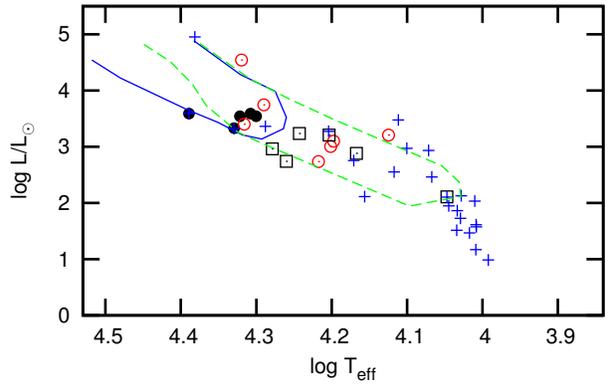}
\caption{The location of SPB/$\beta$~Cep hybrids with spectroscopic parameters 
in  the theoretical H-R diagram are shown as (black) filled circles.
SPB stars with spectroscopic parameters are shown as (red) open circles.
The (black) open squares are anomalous stars with high frequencies which we
call MAIA variables.  Stars which do not pulsate (i.e. are classified as 
neither SPB, BCEP or MAIA) are shown by (blue) plus signs.  The region 
enclosed by the solid line is the theoretical instability strip for 
$\beta$~Cep  variables.  The dashed region is the theoretical instability 
strip for SPB variables.}
\label{hrstrip}
\end{figure}

There are a few stars with reliable (spectroscopic) parameters.  Among these
are five SPB/$\beta$~Cep hybrids (including the Be/BCEP star KIC\,8057661),
six Maia variables and seven SPB stars.  Their location in the H-R diagram is 
shown in Fig.\,\ref{hrstrip}.  Apart from the problematic Maia variables,
there are three SPB stars within the $\beta$~Cep instability strip which show 
no signs of high-frequency modes.  In addition, there are several stars with  
well-determined parameters which show no clear signs of $\beta$~Cep or SPB 
pulsations, but lie within the instability strips.

Another strange aspect of these $\beta$~Cep stars is that all of them have
rich low-frequency spectra.  In fact, the low frequencies are dominant even
in the hottest stars.  Current models of $\beta$~Cep stars have difficulty
in accounting for these hot stars with low frequencies (see below).

\section{The K2-0 data}

Spacecraft drift is the main component leading to increased photometric
scatter compared to the original {\it Kepler} field.  Using simple aperture
photometry (SAP), the brightness of a target changes slowly as the star
moves from its original location on the CCD until the thrusters are
applied to bring the star back to its original location.  The periodogram
therefore shows strong peaks at about 4.08\,d$^{-1}$ and its harmonics
which corresponds to the frequency at which the thrusters are applied.  It
is possible to obtain a greatly improved light curve simply by locating
these peaks in the periodogram and fitting and removing a truncated Fourier 
series.  Extraction of the light curve from the FITS files using SAP, fitting 
and removing the truncated Fourier series was done using the {\sc keplc} 
software.
  
A more sophisticated approach to correcting the light curve has been proposed 
by \citet{Vanderburg2014}.  In this technique the nonuniform pixel response 
function of the {\it Kepler} detectors are determined by correlating flux 
measurements with the spacecraft's pointing and removing the dependence. 
This leads to an improvement over raw SAP photometry by factors of 2--5, with 
noise properties qualitatively similar to {\it Kepler} targets at the same 
magnitudes.  There is evidence that the improvement in photometric precision 
depends on each target's position in the {\it Kepler} field of view, with worst 
precision near the edges of the field. Overall, this technique restores the 
median-attainable photometric precision within a factor of two of the original
Kepler photometric precision for targets ranging from 10--15 magnitude 
in the {\it Kepler} bandpass.  This technique is implemented in the {\sc
kepsff} task within the {\sc PYKE} software suite.

\begin{center}
\begin{table*}
\caption{List of stars in the {\it Kepler} K2-0 field which we
classified as possible $\beta$~Cep and SPB stars.}
\label{bcepspb}
\begin{tabular}{rllrrrrl}
\hline
\multicolumn{1}{c}{EPIC} & 
\multicolumn{1}{c}{Type} & 
\multicolumn{1}{c}{Name} & 
\multicolumn{1}{c}{V} & 
\multicolumn{1}{c}{B-V} & 
\multicolumn{1}{c}{U-B} & 
\multicolumn{1}{c}{$T_{\rm eff}$} & 
\multicolumn{1}{c}{sptype} \\
\hline
\\
 202061127 &   BCEP?     &  2MASS J06041476+2404024  &    14.000 &      0.470 &          &          &  B5np                            \\  
 202061129 &   MAIA      &  Cl* NGC 2129 CCP 32      &    14.068 &      0.672 &    0.341 &    9600  &  B9                              \\  
 202061131 &   MAIA      &  Cl* NGC 2129 CCP 33      &    14.234 &      0.718 &    0.056 &   14200  &  B5                              \\  
 202061216 &   BCEP      &  Cl* NGC 2175 H 43        &    10.580 &      0.220 &          &          &  B2.5\,III-IV                      \\  
 202061002 &   MAIA?     &  GSC 01326-01308          &    12.400 &      0.600 &          &          &  B9                              \\  
 202060712 &   MAIA?     &  HD 251584                &    11.120 &      0.440 &          &          &  B9                              \\  
 202062487 &   BCEP?     &  HD 252945                &    10.260 &      0.040 &          &          &  Om...                           \\  
 202060562 &   MAIA      &  HD 253107                &    10.370 &      0.040 &   -0.590 &   17600  &  B9                              \\  
 202060461 &   BCEP?     &  HD 254700                &     9.830 &      0.330 &   -0.440 &   19400  &  B5                              \\  
 202062129 &   BCEP/MAIA &  TYC 1877-374-1           &    11.620 &      0.430 &   -0.310 &   18000  &  B... (F8\,V)                      \\  
 202062527 &   BCEP?     &  TYC 1881-990-1           &    13.160 &     -0.920 &          &          &  O...                            \\  
 202062107 &   SPB       &  BD+23 1302               &    10.200 &      0.290 &   -0.480 &   19600  &                                  \\  
 202060496 &   SPB?      &  Cl* NGC 2129 PP 30       &    12.886 &      0.814 &          &          &  B4                              \\  
 202061247 &   SPB       &  Cl* NGC 2175 H 33        &    12.690 &      0.320 &          &          &  B5\,III-IV                        \\  
 202060831 &   SPB       &  HD 252532                &    11.680 &     -0.160 &          &          &  B5                              \\  
 202060707 &   SPB       &  HD 252588                &    11.110 &      0.210 &          &          &  B9                              \\  
 202060130 &   SPB?      &  HD 256149                &     9.400 &      0.190 &          &          &  B8V                             \\  
 202062234 &   SPB       &  NSV 2805                 &    11.760 &      0.110 &          &          &  B..                             \\  
 202061048 &   SPB/ROT   &  Cl* NGC 2129 PP 5        &    12.766 &      0.634 &          &          &  B5                              \\  
 202061244 &   SPB/ROT   &  Cl* NGC 2175 H 106       &    12.150 &      0.120 &          &          &  B4.5\,III-IV                      \\  
 202061242 &   SPB/ROT   &  Cl* NGC 2175 H 166       &    12.010 &      0.260 &          &          &  B4\,IV                            \\  
 202062514 &   SPB/ROT   &  GSC 01864-01382          &    12.300 &      0.700 &          &          &  Om...                           \\  
 202061287 &   SPB/ROT   &  HD 253215                &    10.760 &     -0.080 &          &          &  Be                              \\  
 202062078 &   SPB/ROT   &  HD 257387                &     9.870 &      0.330 &          &          &  F0                              \\  
 202062226 &   SPB       &  TYC 1864-314-1           &    11.950 &      0.240 &   -0.700 &   28400  &  B...                            \\
 202062523 &   SPB       &  TYC 1877-1694-1          &    12.610 &     -0.650 &          &          &  O...                            \\  
\\
\hline
\end{tabular}
\end{table*}
\end{center}

\begin{figure*}
\centering
\includegraphics{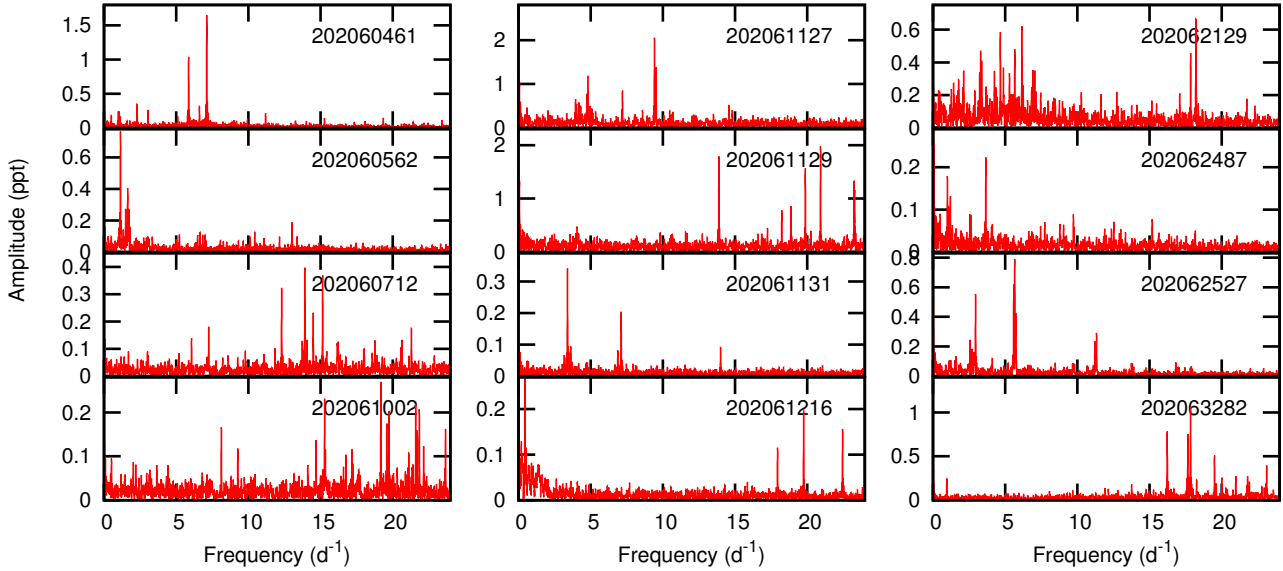}
\caption{Periodograms of candidate $\beta$~Cep stars in the K2-0 field.}
\label{k2bcepr}
\end{figure*}

\begin{figure*}
\centering
\includegraphics{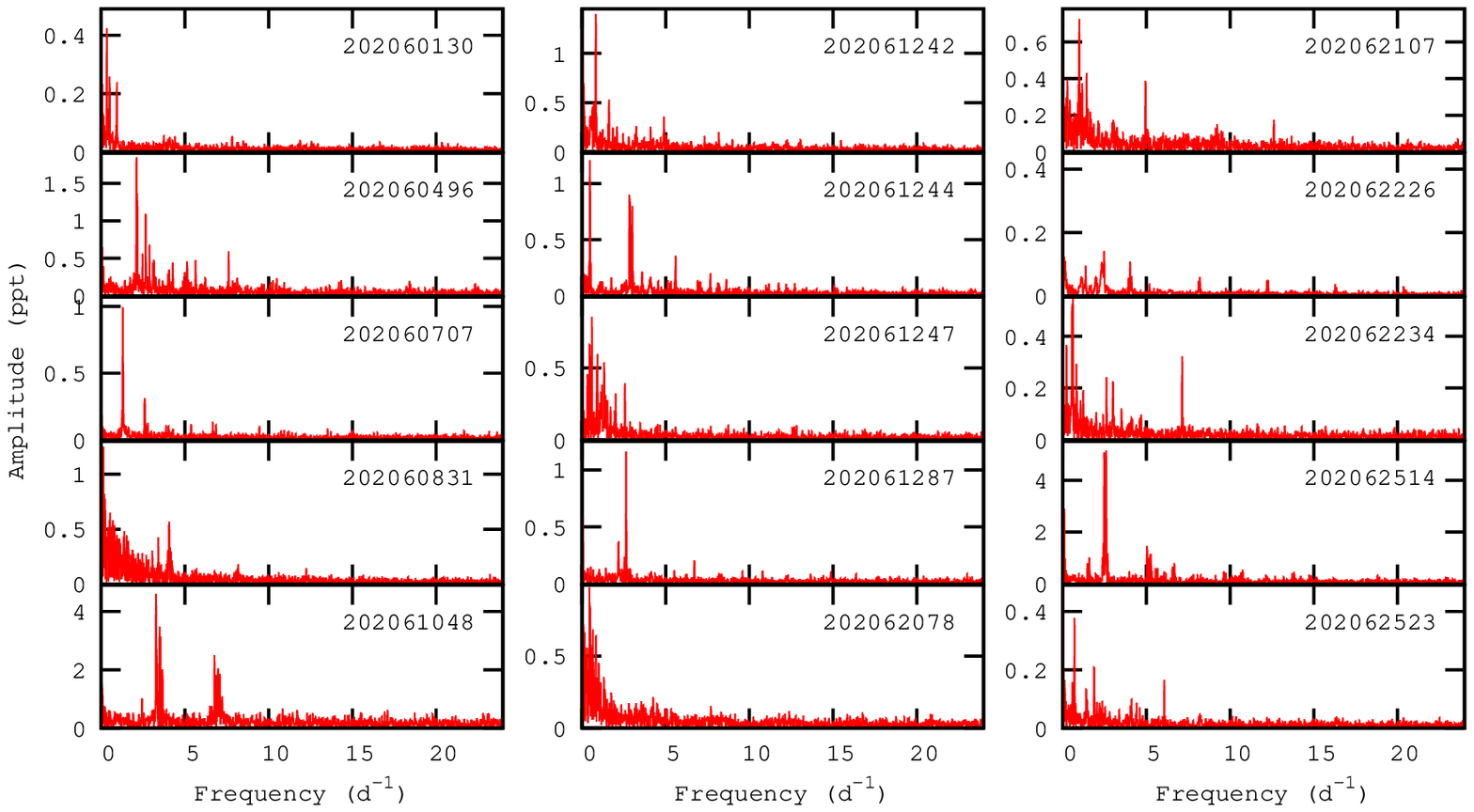}
\caption{Periodograms of candidate SPB stars in the K2-0 field.}
\label{k2spbr}
\end{figure*}

B-type stars in the K2-0 field were selected according to their spectral
types in the {\it Catalogue of Stellar Spectral Classifications}
\citep{Skiff2014}.  As a first task, we examined the light curves and 
periodograms of these B stars extracted by the two methods described above. 
Our aim was to detect interesting stars, particularly $\beta$~Cep and SPB 
variables.  The distinction between main sequence and compact B stars is 
problematic.  In the absence of appropriate spectral classification, we assume 
that bright stars are more likely to be main sequence objects, whereas stars 
with high proper motions are compact objects.  Using these criteria, we have 
classified 300 B stars in the K2-0 field.

Unlike the {\it Kepler} field where ground-based photometry is available for
all stars, very few ground-based observations are available for stars in the
K2-0 field.  We searched for UBV photometry in the K2-0 field using the 
SIMBAD database.  Unfortunately, U-band photometry is a vital requirement in
order to determine the reddening and hence the intrinsic colour of the star. 
Lack of U-band photometry means that even basic information such as the
effective temperature is lacking for most stars.  Since almost no
spectroscopy or Str\"{o}mgren photometry is available for these stars, there
is no possibility of estimating their luminosities.  

Unlike the {\it Kepler} field, for which 4\,yr of superb photometry is
available for all objects, the data for the K2-0 field spans only about
80\,d, but only the last 38\,d are truly useful.  The thruster corrections
severely affect the low-frequency range.  As a result, we cannot determine 
whether a star is an ellipsoidal variable unless the amplitude is very large.  
It is also very difficult to know whether a star is an SPB variable because of 
the low-frequency peaks cannot be entirely trusted to be intrinsic to the
star.  High frequencies are relatively unaffected, and the identification of a
star as a $\beta$~Cep variable is quite easy as long as the amplitude is 
sufficiently large.

Table\,\ref{bcepspb} lists those stars which we believe may be $\beta$~Cep 
and SPB variables.  Fig.\,\ref{k2bcepr} shows periodograms of the
$\beta$~Cep variables and Fig,\,\ref{k2spbr} those of the SPB variables.
It should be noted that peaks as high as 20\,d$^{-1}$ can be seen in some of
these stars.  Such high frequencies are not expected in $\beta$~Cep stars
unless they are close to the zero-age main sequence.  It is clearly
imperative to obtain ground-based data to determine the nature of these
stars.  

\begin{figure}
\centering
\includegraphics{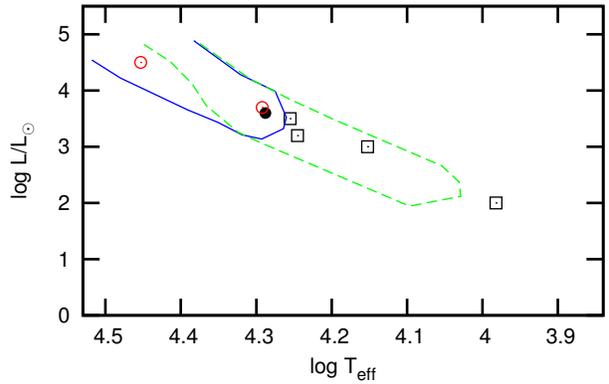}
\caption{The location of K2-0 SPB (red open circles), presumed Maia stars 
(black squares) and $\beta$~Cep variable (black filled circle) in  the 
theoretical H-R diagram.  The luminosities are unknown and the stars have 
been placed near the midpoint of the instability region.  The region 
enclosed by the solid line is the theoretical instability strip for 
$\beta$~Cep  variables.  The dashed region is the theoretical instability 
strip for SPB variables.}
\label{k2hrs}
\end{figure}

The lack of multicolour photometry and spectroscopy for K2-0 stars means 
that it is not possible to place most of the stars in the H-R diagram.
For those few stars where UBV photometry is available, we calculated the
reddening and estimated the effective temperature from the unreddened 
$(B-V)_0$ colour index.  There is insufficient information to estimate the 
luminosities of these stars.  In Fig.\,\ref{k2hrs} we have simply placed 
each star near the center line of the instability strip.

For stars with high frequencies where UBV photometry is not available, we 
cannot tell whether they are $\beta$~Cep or Maia variables.  We therefore
used the available spectral types to decide between the two.  All stars with 
high frequencies and known effective temperatures are cooler than the
red edge of the $\beta$~Cep instability strip and we have classified them
provisionally as MAIA variables.  Further ground-based spectroscopy
or multi-colour photometry is required to exploit the full potential of the
K2-0 data. 

\section{Two challenging problems}

If we assume that the effective temperatures and luminosities derived from
spectroscopy or Str\"{o}mgren photometry are reasonably accurate, we are
faced with two problems which challenge our current understanding
of pulsation in hot stars.  These are: (i) stars in the $\beta$~Cep instability 
strip which exhibit low-frequency peaks and no high frequency peaks, and
(ii) stars in the SPB instability strip with high frequencies typical of
$\beta$~Cep stars (what we refer to as MAIA variables in this paper).

It is possible that a low metallicity, $Z$, may be responsible for (i).  It is 
well known (e.g., \citealt{Pamyatnykh1999}) that in B-type main sequence models, 
g modes of high radial order can remain unstable at much lower values of $Z$ 
than p-modes.  Moreover, reducing the metallicity will shift the location of 
the star in the H-R diagram to the position occupied by a star with solar 
abundance and lower mass.  Also, an increasing number of low-frequency g 
modes become unstable with decreasing mass.

\begin{figure}
\centering
\includegraphics{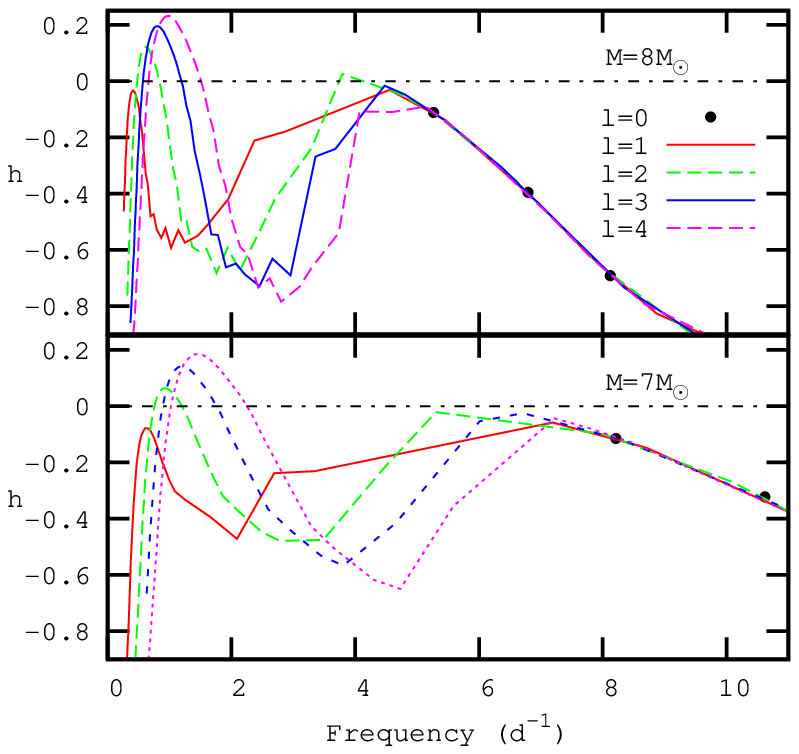}
\caption{The pulsational instability parameter, $\eta$, for modes of low 
degree, $l$, as a function of frequency (in the observer's frame) for a 
model with mass $M = 8M_\odot$, $\log T_{\rm eff} = 4.2974$, 
$\log L/L_\odot = 3.744$  (top panel) and $M = 7M_\odot$, 
$\log T_{\rm eff} = 4.3018$, $\log L/L_\odot = 3.459$ (bottom panel).  
OP opacities and a metallicity $Z = 0.01$ are used.}
\label{lowfrmod}
\end{figure}

In Fig.\,\ref{lowfrmod}, we show the instability parameter, $\eta$, as a 
function of frequency for two models corresponding approximately to the two 
open circle in the $\beta$ Cep instability strip in Fig.\,{k2hrs}.  These 
models were computed with OP opacities, $Z = 0.01$ and the initial hydrogen 
abundance $X_0 = 0.7$.   Modes with harmonic degree $l$ = 0--4 are considered.
The model with $M = 8\,M_\odot$ is close to the end of core hydrogen burning, 
while the model with $M = 7\,M_\odot$ is more or less in the middle of the 
main sequence evolutionary stage.  As can be seen, the low-frequency modes in 
these models have values of $\eta$ comparable or higher than the high-frequency 
modes.  In other words, the low-frequency modes are unstable, while the
high-frequency modes remain stable (though on the point of instability).
Furthermore, the larger the spherical harmonic, $l$, the greater the
instability of low-frequency modes.  

\begin{figure}
\centering
\includegraphics{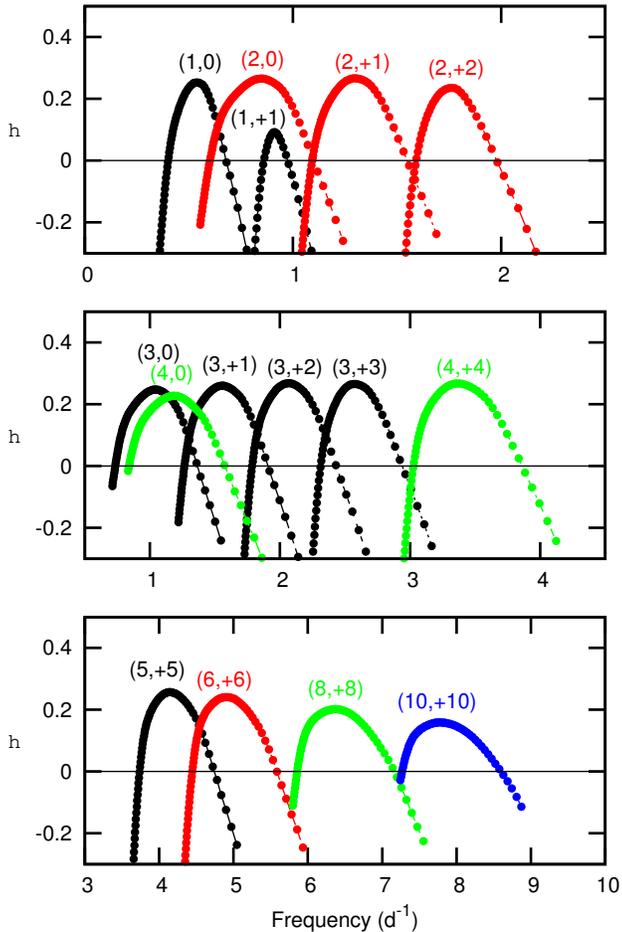}
\caption{The pulsational instability parameter, $\eta$, as a function of frequency for
models with mass $M = 6M_\odot$, $\log T_{\rm eff} = 4.1939$, $\log L/L_\odot 
= 3.299$ and equatorial rotational velocity of 200\,km\,s$^{-1}$.}
\label{trad}
\end{figure}

For these models, the low frequencies can be easily understood, but high 
frequencies cannot be driven.  In fact, it is very difficult to drive 
high-frequency, $\beta Cep$-like modes in stars with masses $M < 6\,M_\odot$.  
With our current understanding of pulsation in B-type main sequence stars, 
the only explanation seems to be a shift of low-frequency modes to higher 
frequencies due to rotation.  In Fig.\,\ref{trad}, we show the instability 
parameter, $\eta$, as a function of frequency for a model with $6\,M_\odot$ 
and an equatorial rotational velocity $v = 200$\,km\,s$^{-1}$. Modes with 
$(l, m)$ up to (10,+10) were considered.  In the pulsational computations, 
the effects of the Coriolis force were taken into account in the framework 
of the traditional approximation (e.g., \citealt{Townsend2003a, Townsend2005, 
Jadwiga2007}).  We found unstable modes for masses as low as $M \approx 
3.5\,M_\odot$.  Note that prograde modes with $l = 10$,  $m = +10$ can 
attain a frequency of about 8.5\,$d^{-1}$.  Some SPB stars in our sample have 
high values of $v \sin i$.  For example, KIC\,12258330  rotates with an 
equatorial velocity in excess of $v > 130$\,km\,s$^{-1}$ and KIC\,10960750 
with $v > 250$\,km\,s$^{-1}$.  This might be the explanation for what we call 
MAIA variables in this paper.  However, it is very difficult to understand 
frequencies higher than about 10\,d$^{-1}$, as found in some of these stars.

If a rotational shift of low frequencies to high frequencies is responsible for 
the high frequencies in stars with $M < 6M_\odot$, one may expect an asymmetry 
in driving prograde and retrograde modes.  Such an asymmetry was suggested by
\citet{Lee1989} who used a truncated expansion for the eigenfunctions, but is 
not confirmed by computations with the traditional approximation (e.g.,
\citealt{Dziembowski2007b}).  Finally, it may be that an entirely different 
driving mechanism is responsible for the Maia stars.  This may perhaps be 
related to the opacity bump at  $\log T = 5.1$ discussed by 
\citet{Cugier2012}.  However, it seems to us that the number of stars is too 
small to be compatible with this mechanism.

\section{Discussion}

In \citet{Balona2011b} stars were grouped into different classes depending
on the appearance of the periodogram.  We did not find such a classification
scheme useful, preferring instead to search for physical causes of the
variation (tidal distortion, rotational modulation, high- and low-frequency
pulsations).  Without intensive ground-based observations, particularly
high-resolution spectroscopy, it is not possible to confirm whether our
classifications are correct, but they at least provide a useful starting
point.

Of the 115 stars in the {\it Kepler} field that we examined, 32 (or 
28\,percent) appear to be  ellipsoidal variables.  On the other hand, there
is only one eclipsing binary in the sample.  Ellipsoidal variables must be 
seen at low inclinations for no eclipses to occur.  If the orbital axes are 
randomly orientated the probability of finding a low inclination, $i$, is 
proportional to $\sin i$.  Because these stars have short periods and
therefore close to each other, one expects to see many more eclipsing 
systems than ellipsoidal variables.  The probability of observing 28\,percent
ellipsoidal variables but less than one percent eclipsing binaries in a
random group of stars must therefore be very small.  Perhaps the most 
plausible explanation is that the variations are not orbital in nature but a 
result of stable, co-rotating features.  The origin of these variations must 
remain speculative until such time that high-resolution time series 
spectroscopy can be obtained.

In the past, the idea of rotational modulation in B stars has usually been
considered improbable because of the perceived notion that radiative
envelopes cannot support star spots or other surface inhomogeneities.  The
discovery that about 40\,percent of {\it Kepler} A stars do indeed have
spots with periods compatible with the known distribution of rotation periods 
of A stars \citep{Balona2013c}, shows that spots or co-rotating structures 
must be considered in early-type atmospheres.  The large number of stars 
classified as rotational variables (type ROT; 45 stars or 39\,percent) is 
testimony that spots or co-rotating structures exist in a large fraction of B 
stars.

We find nine examples (8\,percent) of the strange ROT-d pattern 
(Fig.\,\ref{mess}) which is so common among A stars \citep{Balona2013c, 
Balona2014b}. Most are late B stars, but one of them, KIC\,5458880, is 
classified as B0\,III.  A satisfactory explanation for the periodogram 
structure has not been found, but it is probably related to rotation.
A careful study of rotational modulation using high-dispersion spectroscopy 
would surely lead to a further understanding of the atmospheres of B stars.

These surprising conclusions open up a new perspective on B star
atmospheres.  The fact that so many B stars appear to show rotational
modulation suggests that the role of magnetic fields and activity in B star
atmospheres has been underestimated in the past.  Perhaps star spots may be
a result of a magnetic field generated by the Tayler instability in a 
differentially rotating star \citep{Spruit2002, Mullan2005}.  

A particularly puzzling aspect, which was already mentioned in
\citet{Balona2011b}, is that $\beta$~Cep pulsations are always accompanied
by a rich spectrum of low-frequency modes.   One possibility is that these 
might be stars of low metallicity where low-frequency g modes tend to be more 
unstable, relative to the high-frequency modes, than in stars with higher 
metallicity.  For the moment, however, the problem is unresolved.

Another puzzle, already mentioned in \citet{Balona2011b}, is the incidence of 
high-frequency pulsations in stars which are cooler than the red edge of the 
$\beta$~Cep instability strip.  It has been assumed that these may be binaries
in which a $\delta$~Sct secondary companion is in orbit around a more luminous
B-type primary.  This would be an acceptable explanation if the numbers of
such stars were few.  The fact more than half of the sample needs to be 
explained in this way makes this explanation very unlikely.  The fact that 
similar stars have been observed by {\it CoRoT} \citep{Degroote2009b} and 
also in the K2-0 field as well as from ground-based observations of stars in a
young open cluster \citep{Mowlavi2013}, suggests that another explanation is 
required.  We have therefore provisionally assumed that these are examples
of the so-called Maia variables.  There are pulsating stars situated between 
the red edge of the $\beta$~Cep instability strip and the blue edge of the 
$\delta$~Sct instability strip.  

This does not necessarily mean that a new driving mechanism needs to be found
for Maia variables.  It is perhaps possible that the high frequencies are 
low-frequency g modes generally associated with SPB stars shifted to high 
frequencies by rotation. In fact, it has long been a puzzle why nearly all 
SPB stars observed from the ground are slow rotators.  The resolution to this 
problem may be just the Maia variables discussed here.  In other words, the 
SPB stars have the same rotational velocity distribution as normal B stars, 
but it is only the low frequencies which attain sufficient amplitudes to be 
easily detected.  However, frequencies higher than about 10\,d$^{-1}$, as 
found in some Maia stars, cannot be easily understood in this way.

\section*{Acknowledgments} 

This paper includes data collected by the {\it Kepler} mission. Funding for the
{\it Kepler} mission is provided by the NASA Science Mission directorate.
The authors wish to thank the {\it Kepler} team for their generosity in
allowing the data to be released and for their outstanding efforts which have
made these results possible.  

Much of the data presented in this paper were obtained from the
Mikulski Archive for Space Telescopes (MAST). STScI is operated by the
Association of Universities for Research in Astronomy, Inc., under NASA
contract NAS5-26555. Support for MAST for non-HST data is provided by the
NASA Office of Space Science via grant NNX09AF08G and by other grants and
contracts.
 
LAB wishes to thank the South African Astronomical Observatory and the
National Research Foundation for financial support.

\bibliographystyle{mn2e}
\bibliography{bstars}

\label{lastpage}

\end{document}